\documentclass[fleqn,usenatbib]{mnras}

\usepackage{newtxtext,newtxmath}
\usepackage{xspace}
\usepackage{xcolor}
\usepackage{threeparttable}

\usepackage[T1]{fontenc}

\DeclareRobustCommand{\VAN}[3]{#2}
\let\VANthebibliography\thebibliography
\def\thebibliography{\DeclareRobustCommand{\VAN}[3]{##3}\VANthebibliography}

\usepackage{graphicx}	
\usepackage{amsmath}	
\usepackage{mhchem}

\newcommand{\kcalmol}{\,kcal mol$^{-1}$}
\newcommand{\syn}{{\emph{syn}\xspace}}
\newcommand{\anti}{{\emph{anti}\xspace}}
\newcommand{\rev}[1]{#1}
\newcommand{\revt}[1]{#1}

\title[\syn-Propenethial in the ISM]{Laboratory spectroscopy, theoretical characterization, and astronomical search for syn-\rev{propenethial} (\ce{CH2CHCHS})}

\author[G. Molpeceres et al.]{
Germán Molpeceres,$^{1}$\thanks{E-mail: german.molpeceres@iff.csic.es (GM)}
Carlos Cabezas,$^{1}$\thanks{E-mail: carlos.cabezas@csic.es (CC)}
Marcelino Agúndez,$^{1}$
María Mallo,$^{1}$
Yasuki Endo,$^{2}$
\newauthor
Lucie Kolesnikov\'a,$^{3}$
Gisela Esplugues,$^{4,5}$
José Cernicharo$^{1}$
\\
$^{1}$Instituto de Física Fundamental, Consejo Superior de Investigaciones Científicas, Calle Serrano, 113b, 121, 123, 28006, Madrid, Spain\\
$^{2}$Department of Applied Chemistry, Science Building II, National Yang Ming Chiao Tung University, Hsinchu 300098, Taiwan\\
$^{3}$Department of Analytical Chemistry, University of Chemistry and Technology,Technick\'{a} 5, 166 28 Prague 6, Czechia\\
$^{4}$Observatorio Astronómico Nacional (OAN, IGN), C/ Alfonso XII, 3, 28014 Madrid, Spain
\\
$^{5}$Observatorio de Yebes, IGN, Cerro de la Palera s/n, 19141 Yebes, Guadalajara, Spain
}

\date{Accepted XXX. Received YYY; in original form ZZZ}

\pubyear{2015}

\begin{document}
\label{firstpage}
\pagerange{\pageref{firstpage}--\pageref{lastpage}}
\maketitle

\begin{abstract}
We report the laboratory characterization of the higher energy isomer of propenethial, \syn-\ce{CH2CHCHS}. While the \revt{lower-energy} isomer \anti-\ce{CH2CHCHS} was detected in the interstellar medium in the course of the \textsc{Quijote} line survey of TMC-1, we report the non-detection of the \syn-isomer in the same source, with a derived upper limit to its column density of 1.5$\times$10$^{10}$ cm$^{-2}$. A subsequent theoretical investigation for the causes of the non-detection reveals \revt{that} the most plausible route for the formation of \ce{CH2CHCHS} is isomer specific, with nearly a 95\% of formation of the \anti-isomer. This ratio \revt{allows} us to derive an even lower upper limit for the molecule. Besides, studies of tunneling mediated unimolecular isomerization in the gas reveal that \syn-\ce{CH2CHCHS} converts to the low energy isomer in timescales \revt{of} the order of 10$^{3}$ years. Overall, we conclude that the detection of \syn-\ce{CH2CHCHS} is very challenging on cold and warm environments alike. Our results underscore the importance of electronic and kinetic effects \revt{affecting} isomer abundances in diverse interstellar environments.
\end{abstract}




\section{Introduction}

The last decade has witnessed a spectacular increase in the number of complex organic molecules (COMs) detected in the interstellar medium (ISM), with nearly 350 molecules identified to date.\footnote{Number of molecules taken from the March 2026 update of the CDMS database, \url{https://cdms.astro.uni-koeln.de/classic/molecules}} Among these species there are two families for which detections have been very abundant in the last years: sulphur bearing molecules \citep{2021A&A...646L...3C,2021A&A...648L...3C,  cernicharo_sulphur_2021, 2022A&A...657L...4C, 2024A&A...682L...4C, 2024A&A...688L..13C, 2025A&A...693L..20A, remijan_missing_2025, sanz-novo_abiotic_2025, araki_detection_2026} and high-energy isomers \citep{san_andres_first_2024, cernicharo_discovery_2022, rivilla_first_2023, 2024A&A...682L...4C, cernicharo_discovery_2024, esplugues_discovery_2026, molpeceres_formic_2025, 2025A&A...703A..73Z, sanz-novo_conformational_2025}, to indicate some recent works from the literature.

\begin{figure}
\includegraphics[angle=0,width=\linewidth]{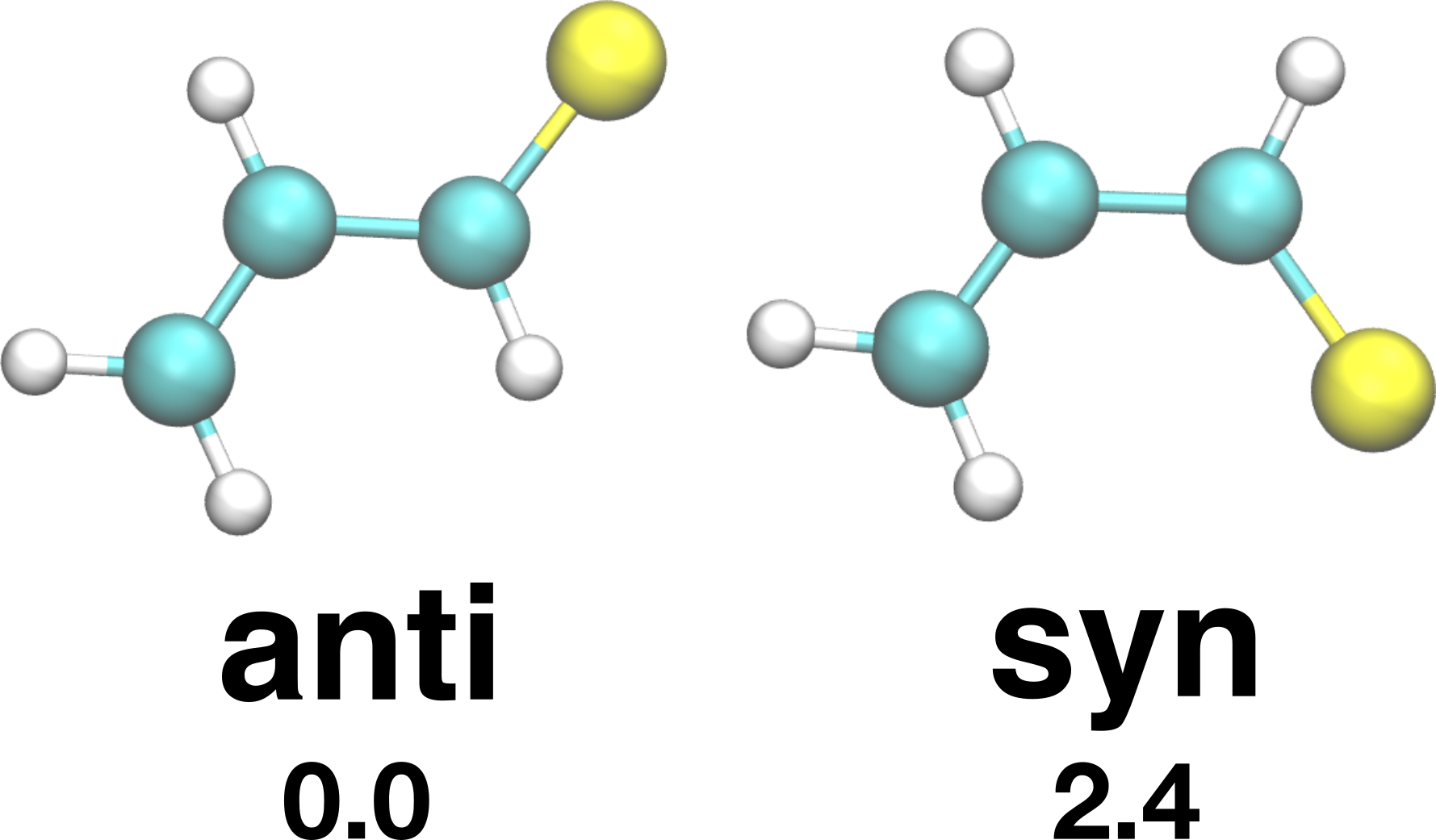}
\label{fig:isomers}
\centering
\caption{Molecular structure of the two \rev{conformational isomers} that are connected through a single bond rotation, \syn\ and \anti\ \ce{CH2CHCHS}. \revt{The numbers represent the zero-point energy corrected energy separation in \kcalmol at the CCSD(T)-F12/cc-pVTZ-F12//SCS-MP2/aug-cc-pVTZ, described later in the text. }} 
\end{figure}

One of the most recent detections of sulphur bearing species reported in the literature is the \anti\ isomer of the propenethial molecule (sometimes also named thioacroleine), \ce{CH2CHCHS} \citep{cabezas_discovery_2025}, in the TMC-1 molecular cloud during the course of the \textsc{Quijote} line survey \citep{Cernicharo2021, cernicharo_quijote_2022}. This molecule, the sulphur analogue of acroleine (\ce{CH2CHCHO}), also detected in TMC-1 \citep{2021A&A...649L...4A}, was found with a column density of 4.4$\pm$0.4$\times$10$^{10}$~cm$^{-2}$, placing it as the least abundant sulphur bearing molecule detected in TMC-1. Propenethial is a molecule with two possible conformational isomers, \syn\ and \anti\ (shown in Figure \ref{fig:isomers}), with the latter being the most stable one by 2.4 \kcalmol (i.e. around 1207 K).\footnote{The numerical values for the energy difference and interconversion barrier between isomers are obtained from our own calculations shown later in the text} The \anti\ isomer was the one detected in TMC-1, while the \syn\ isomer was not searched for due to the unaivailability of its experimental spectroscopic constants, although, during the course of this work theoretical predictions were reported \citep{rauhut_spectroscopic_2026}. The interconversion barrier in both isomers of propenethial is significantly higher (5.3 \kcalmol~in the exothermic direction) than the energy difference between the isomers, suggesting that the \syn\ isomer, if formed, should be stable under TMC-1 conditions. The presence of high energy isomers in the ISM is a precious tool to not only constrain the chemical formation routes of the molecules, but also as a probe of the physical conditions where these molecules are formed \citep{hacar_hcn--hnc_2020, loison_interstellar_2017, 2021ApJ...923...24H, rivilla_interstellar_2026}. 

Another relevant aspect of the detection of propenethial concerns the anomalous abundance ratios between aldehydes and their thione analogues, -C(H)=S. One of the longstanding puzzles in astrochemistry is the so called missing sulphur problem \citep{1999MNRAS.306..691R, vidal_reservoir_2017}. Although it is now clear that the sulphur bearing molecules detected so far in TMC-1 cannot account for the cosmic sulphur abundance by a factor of two orders of magnitude, a direct comparison with their oxygen bearing counterparts provides a useful way to disentangle the specific contribution of each chemical family. In \citet{cabezas_discovery_2025} we showed that the abundance ratios between aldehydes and thiones are highly scattered. The observed values are 4.7 for \ce{CH2CHCHS}/\ce{CH2CHCHO}, 35 for \ce{CH3CHO}/\ce{CH3CHS} \citep{2025A&A...693L..20A}, 46 for HCCCHO/HCCCHS \citep{loison_interstellar_2016, cernicharo_sulphur_2021}, and 0.27 for NCCHO/NCCHS \citep{2024A&A...688L..13C}. While the dispersion in these ratios is evident, the correct values could be affected by the presence of high energy isomers. Just to give another example, the \syn\ and \anti\ isomers of crotononitrile, \ce{CH3CHCHCN}, have been observed with comparable abundances, in contrast with the thermodynamic ratio expected from their relative energies \citep{cernicharo_discovery_2022, mallo_isomer-specific_2025}.

In this work, we have investigated the rotational spectrum of the \syn\ isomer of propenethial, obtaining a set of accurate rotational and distortion constants that allowed us to search for this species in TMC-1. As we show later in the text (Section \ref{sec:observations_results}), our search proved unsuccessful. The non-detection of some high-energy isomers, contrary to the detection of others \citep{cernicharo_discovery_2022, 2024A&A...682L...4C,cernicharo_discovery_2024,molpeceres_formic_2025, esplugues_discovery_2026, remijan_missing_2025,rivilla_interstellar_2026} provides valuable insight into the puzzle that interstellar isomerism poses, particularly in chemically rich regions like TMC-1. Therefore, in this work we also delve into the quantum chemical investigation of the reasons for the absence of \syn-propenethial. Our work is structured as follows. In Section \ref{sec:methods} we describe all the methodologies used in this work, including laboratory experiments, astronomical observations and quantum chemical calculations, followed by a detailed description of the results in Section \ref{sec:results}. We combine all the results attained in the previous sections to provide a wider picture of the place propenethial isomers occupy in the chemical landscape of the ISM and its potential detectability. Finally, we provide a small summary of our findings in Section \ref{sec:conclusions}. 

\section{Methodology} \label{sec:methods}

\subsection{Laboratory experiments}

In our previous work \citep{cabezas_discovery_2025}, we measured the rotational spectrum of propenethial in the 144--179 and 293--321~GHz frequency regions using the Prague semiconductor millimeter wave spectrometer \citep{Kania2006}. In this experiment, propenethial molecules were generated in the gas phase by pyrolysis of diallyl sulfide (\ce{\ce{(CH2CHCH2)2S}}) at the temperature of about 500~$^{\circ}$C. The recorded spectrum was significantly congested primarily due to  transitions arising from the ground vibrational state and numerous vibrationally excited states of the \anti\ isomer. Consequently, it was not feasible at that time to identify any lines attributable to the propenethial \syn\ isomer, since its lines are expected to be approximately 90–100~times weaker than those of the \anti\ isomer.

In this work, we employed Balle-Flygare narrowband-type FTMW spectrometer with an electric discharge source \citep{Endo1994,Cabezas2016a} to observe the rotational transitions of the \syn\ isomer of propenethial. The use of this FTMW technique has some advantages over millimeter wave spectrometer employed in the previous study in terms of observing the \syn\ isomer of propenethial. First, since the former is an experiment done in a supersonic expansion, only the rotational levels of the ground vibrational state will be populated, which significantly simplifies the observed spectrum. Secondly, the use of a highly energetic electric discharge in the generation of propenethial can increase the abundance of the \syn\ isomer versus the \anti\ isomer, due to the high kinetic temperature of the discharge.

In the present experiment, propenethial molecules were produced by pulsed electric discharge in a supersonic jet of a gas mixture of diallyl sulfide diluted in Ar. To prepare the chemical mixture, a small amount of diallyl sulfide sample (Sigma-Aldrich, ~97\%, b.p.~ 138~$^\circ$C) was placed in a liquid reservoir placed immediately before the pulsed valve. A flow of Ar gas at a pressure of 1.0~bar was passed through a liquid container. This highly diluted mixture of diallyl sulfide in Ar flowed through the pulsed-solenoid valve that is accommodated in the backside of one of the cavity mirrors and aligned parallel to the optical axis of the resonator. A pulse voltage of 900~V with a duration of 450~$\mu$s was applied between stainless-steel electrodes attached to the exit of the pulsed discharge nozzle, resulting in an electric discharge synchronized with the gas expansion. The products generated in the discharge were then rapidly cooled to a rotational temperature of $\sim$2~K and were probed by Fourier-transform microwave spectroscopy in the 4-40~GHz region. The estimated accuracy of the frequency measurements is better than 3~kHz, and the resolution better than 5~kHz.

\subsection{Observations}

The molecule \syn-\ce{CH2CHCHS} was searched for in the cold dense cloud Taurus Molecular Cloud 1 (TMC-1). For that purpose we used the \textsc{Quijote}\footnote{Q-band Ultrasensitive Inspection Journey to the Obscure TMC-1 Environment} dataset \citep{Cernicharo2021}. Briefly, \textsc{Quijote} consists of an unbiased spectral line survey carried out in the Q band (31.0-50.3 GHz) with the Yebes\,40m telescope in the frame of the \textsc{Nanocosmos} project\footnote{ERC-2013-Syg-610256-NANOCOSMOS (\url{https://nanocosmos.iff.csic.es})} at the position of the cyanopolyyne peak of TMC-1 ($\alpha_{J2000}=4^{\rm h} 41^{\rm  m} 41.9^{\rm s}$ and $\delta_{J2000}=+25^\circ 41' 27.0''$). A 7\,mm dual linear polarization receiver was used connected to a set of eight fast Fourier transform spectrometers. This setup permits to observe the full Q band in both polarizations in one shot with a spectral resolution of 38.15 kHz. The system is described in detail in \cite{Tercero2021}. The observations of \textsc{Quijote} were carried out in the frequency-switching observing mode, with a frequency throw of either 8 MHz or 10 MHz. The latest \textsc{Quijote} dataset includes observations carried out between November 2019 and July 2024, with a total on-source telescope time of 1509.2 h. The intensity scale is the antenna temperature, $T_A^*$, which has a calibration uncertainty of 10\,\%. The $T_A^*$ rms noise level varies between 0.06 mK at 32 GHz and 0.18 mK at 49.5 GHz. The procedure used to reduce and analyze the data was described in \cite{Cernicharo2022}. All data were analyzed using the \textsc{Gildas} software.\footnote{\url{https://www.iram.fr/IRAMFR/GILDAS/}}

\subsection{Quantum chemical calculations}

\subsubsection{Spectroscopy} \label{sec:spec_qc}

We performed quantum chemical calculations to optimize the molecular structure of the \syn-\ce{CH2CHCHS} to obtain precise values of the rotational constants, in order to search for the rotational transitions of this species in the spectrum. For the geometry optimization, we used the coupled cluster method with single, double, and perturbative triple excitations (CCSD(T)) method and an explicitly correlated approximation (F12A; \citealt{Knizia2009}). The Dunning's correlation consistent polarized valence triple-$\zeta$ basis set for explicitly correlated calculations (cc-pVTZ-F12; \citealt{peterson_systematically_2008}) was employed. At the optimized geometry, electric dipole moment components were calculated at the same level of theory used for geometrical optimization. All these calculations were performed using the \textsc{Molpro} program \citep{Molpro2020}. The calculated values for the rotational constants as well as the electric dipole moment components derived from these calculations are shown in Table \ref{constants}. In our previous work \citep{cabezas_discovery_2025}, we already reported theoretical values for the rotational constants of \syn-\ce{CH2CHCHS} at the CCSD/cc-pVTZ level of theory. They differ only slightly from those reported in this work.

\subsubsection{Reactivity} \label{sec:gas}

We carried out dedicated calculations to elucidate the potential presence of \syn-\ce{CH2CHCHS} in TMC-1. We first investigated the gas phase chemistry of the most favorable route for the formation of \ce{CH2CHCHS} in \citet{cabezas_discovery_2025}. In our previous theoretical characterization, after scouting possible formation routes, \revt{we} determined that the most promising reaction pathways for the formation of \ce{CH3CHCHS}, independent of the isomer were the \ce{CH2CHCH2 + S -> CH2CHCHS + H} gas-phase reaction and the \ce{C2H3 + HCS -> CH2CHCHS} surface reaction (Appendix B of \citealt{cabezas_discovery_2025}). In this work, we focus in the gas-phase reaction route. All electronic structure calculations are carried out using the \textsc{Orca} code (v.6.1.0) \citep{neese_software_2025}. Additionally we also determined the lifetimes of isomerization of both isomers to determine, \revt{if} \syn-propenethial is formed, to which extent is stable under high-vacuum conditions.

\paragraph*{Gas-phase formation.} We investigated the \rev{reaction of propenyl radical (misidentified in \citet{cabezas_discovery_2025} as propyl) with atomic sulfur} (\ce{CH2CHCH2 + S -> CH2CHCHS + H})  using a simplified framework to evaluate the potential isomer selective nature of the reaction. \rev{The present treatment assumes that alternative neutral-neutral and ion-molecule formation routes play a secondary role under the conditions considered, and therefore constitutes an approximation, in absence of more data.} The simplified kinetic landscape assumes capture of the sulphur atom on the sides of the propenyl (\ce{CH2CHCH2}) radical, that, as we report in \citet{cabezas_discovery_2025} is not the only attack position of the molecule, as sulphur can also insert in the central position forming the cyclic \ce{CH2CH(S)CH2} adduct. Such a channel will not lead to \syn-\anti\ propenethial imbalances, as these will arise exclusively from the adduct formed after the addition on the side position. Because we are not directly interested in the total value of the rate constants but rather in the relative branching between the \syn\ and \anti, the omission of alternative channels in the reaction is justified. It is also worth noting that, in \citet{cabezas_discovery_2025}, we found that the reaction \ce{C2H3 + H2CS} proceeds via a small barrier of 0.7~\kcalmol, and is therefore unlikely to compete efficiently with the reaction investigated here. Although this barrier is low and within the uncertainty of the theoretical method employed, we note that the adduct formed in this reaction (see Table~B.1 of \citealt{cabezas_discovery_2025}) is the same as that formed in the \ce{CH2CHCH2 + S} reaction. Consequently, assuming that the reaction leads to \ce{CH2CHCHS}, the branching ratio between the \textit{syn} and \textit{anti} isomers is expected to be the same in both cases.

We constructed a simplified potential energy surface (PES) for the \ce{C3H5 + S -> CH2CHSH + H} reaction determining the stationary points of it at the SCS-MP2/aug-cc-pVTZ \citep{grimme_spincomponentscaled_2012, Woon1994} level of theory, using an unrestricted Hartree-Fock wavefunction with energies refined from a CCSD(T)-F12/cc-pVTZ-F12 energy calculation \citep{purvis_full_1982, peterson_systematically_2008,pavosevic_geminal-spanning_2014}. This level of theory, used also for the remaining calculations explained in this section, should provide energies close or below chemical accuracy, and therefore errors below 1 \kcalmol\xspace are expected for the remaining of the manuscript. The protocol to characterize the stationary points is the standard of theoretical reaction kinetics consisting on a complete force minimization and molecular Hessian evaluation to confirm the nature of the stationary points, labelling the corresponding minima and transition states on the surface. 

The relative branching ratio of the \syn\ and \anti\ isomers from the studied reaction is obtained from the ratio in their theoretically derived rate constants, that we determine using a master equation formalism where the transitions between the different states (the stationary points in the PES) are modelled through RRKM theory. Again, because we are interested in relative branching ratios of the \ce{CH2CHCHS} channel, an accurate modelling of the entrance channel is unnecessary as errors in that part will equally affect \syn\ and \anti\ isomers. Therefore, we assume an $r^{-6}$ dependence on the interfragment distance between S and \ce{C3H5} for the capture process, and model it using phase-space theory with an estimated $C_{6}$ coefficient of 100~a.u. The master equation simulations were performed with the \textsc{MESS} code \citep{Georgievskii2013}.

The computed phenomenological rate constants in each of the reaction channels is subsequently used to formulate the relative branching ratios as:

\begin{equation}
	BR = \dfrac{k_{\rm X}}{k_{\rm syn} + k_{\rm anti}}
\end{equation}
where X is the isomer of interest and $k$ the rate constants.

\paragraph*{Calculation of lifetimes using small curvature tunneling calculations.}

We also estimate the temperature dependent lifetimes for the different isomers considering exclusively the spontaneous isomerization \syn\ce{<=>}\anti\ in equilibrium, similar to our recent works \citep{GarciadelaConcepcion2022, molpeceres_formic_2025, del_valle_atom_2026}. In particular, in this work we follow the methodology shown in \citet{GarciadelaConcepcion2022}, in which the unimolecular (thermal) rate constant is defined as:

\begin{equation} \label{eq:sct-cvt}
k_{\rm uni} = \kappa \Gamma \frac{k_{\rm B}T}{h}
\frac{\mathcal{Q}_{\rm TS}}{\mathcal{Q}_{\rm RS}}
\exp\left(\frac{-E^{\ddagger}}{k_{\rm B}T}\right),
\end{equation}
that is the usual expression for canonical transition state theory, where $\mathcal{Q}$ represent the partition function of transition state (TS) and reactant state (RS; either the \syn\ or \anti\ isomers) and $E^{\ddagger}$ is the activation energy, corrected by two factors. The first one, $\Gamma$, accounts for the possibility of recrossing after surpassing the transition state, an effect mediated by thermal fluctuations and neglected in conventional transition state theory. The second one, $\kappa$, is the transmission coefficient for quantum tunneling. In this work $\Gamma$ is obtained according to canonical variational transition state theory (CVTST) \citep{Bao2017} and $\kappa$ is obtained with small-curvature tunneling (SCT) techniques \citep{skodje_general_1981}, which are critical to estimate the rate constants at low temperatures considering the torsion involving the lightweight hydrogen atom. We also use the quantized reactant state approximation \citep{wonchoba_surface_1995} at low temperatures for the transition mode, which allows to abbreviate the totality of our theoretical methods as SCT-CVTST. We use the \textsc{Pilgrim} code in our rate constant evaluation \citep{ferro-costas_pilgrim_2020}. The electronic structure solver for the evaluation of the rate constants is the exact same one as in the study of chemical reactivity, CCSD(T)-F12/cc-pVTZ-F12//SCS-MP2/aug-cc-pVTZ. The CCSD(T)-F12/cc-pVTZ-F12 energy refinement of the minimum energy path (MEP) is carried out by interpolating along 20 points (10 in either direction) of the MP2 MEP using the ISPE algorithm \citep{chuang_mapped_1999}. The lifetimes for \syn\ce{<=>}\anti\ conversion can then be estimated as $\tau$=$\ln2$/$k_{\rm uni, X,Y}$. It must be stressed that these lifetimes are estimated based solely on the unimolecular isomerization process and any other chemical event destroying \ce{CH2CHCHS} (independent of the isomer) should have completely different timescales and might dominate the overall lifetime of the molecule in the ISM.

\section{Results} \label{sec:results}

\subsection{Laboratory characterization}

\begin{table*}
\begin{center}
\caption[]{Spectroscopic constants of \syn-propenethial ($A$-reduction,  I$^{\text{r}}$-representation)$^a$ in its ground vibrational state and for its $^{34}$S isotopic species. }
\label{constants}
\begin{tabular}{lccccc}
\hline
\hline
 & FTMW  &   FTMW-$^{34}$S    & mmW  & FTMW+mmW   &  ab initio$^b$  \\
\hline
$A           $ /MHz                         &   17778.90983(43)  &        17699.74(53) &   17778.922(28)      &    17778.91019(36)      &    17770            \\                                                                                                                                                                                                                                                                                                                                                                                            
$B           $ /MHz                         &    3812.15052(38)  &      3716.07750(56) &    3812.15006(42)    &     3812.150299(36)     &    3820             \\                                                                                                                                                                                                                                                                                                                                                                                            
$C           $ /MHz                         &    3140.60125(37)  &      3072.72200(54) &    3140.60072(29)    &     3140.601514(36)     &    3144             \\                                                                                                                                                                                                                                                                                                                                                                                            
$\Delta_{J}  $ /kHz                         &       1.7494(24)   &         1.665(21)   &     1.748080(87)     &      1.748217(38)       &    1.677            \\                                                                                                                                                                                                                                                                                                                                                                                            
$\Delta_{JK} $ /kHz                         &    --11.650(20)    &        ---11.01(66) &     -11.6238(14)     &     --11.6211(11)       &   --11.423          \\                                                                                                                                                                                                                                                                                                                                                                                            
$\Delta_{K}  $ /kHz                         &      67.78(15)     &      67.789$^c$     &      67.97(42)       &       67.789(91)        &    66.107           \\                                                                                                                                                                                                                                                                                                                                                                                                         
$\delta_{J}  $ /kHz                         &       0.4310(11)   &       0.431742$^c$  &       0.431758(45)   &        0.431742(14)     &    0.411            \\                                                                                                                                                                                                                                                                                                                                                                                                         
$\delta_{K}  $ /kHz                         &       5.67(17)     &       5.5228$^c$    &       5.5245(62)     &        5.5228(34)       &    5.040            \\                                                                                                                                                                                                                                                                                                                                                                                                         
$\Phi_{J}    $ /mHz                         &       ...          &       ...           &       [ 0.43]$^d$    &        [ 0.43]          &     0.43            \\                                                                                                                                                                                                                                                                                                                                                                                                         
$\Phi_{JK}   $ /Hz                          &       ...          &       ...           &         0.02944(36)               &            0.03005(28)  &    0.029            \\                                                                                                                                                                                                                                                                                                                                                                                            
$\Phi_{KJ}   $ /Hz                          &       ...          &       ...           &        --0.3867(29)               &          --0.3858(28)   &   --0.374           \\                                                                                                                                                                                                                                                                                                                                                                                            
$\Phi_{K}    $ /Hz                          &       ...          &       ...           &       [ 1.354]$^d$   &       [ 1.354]$^d$  &    1.354   \\                                                                                                                                                                                                                                                                                                                                                                                                                      
$\phi_{J}    $ /mHz                         &       ...          &       ...           &       [ 0.268]$^d$   &       [ 0.268]$^d$  &    0.268   \\                                                                                                                                                                                                                                                                                                                                                                                                                      
$\phi_{JK}   $ /mHz                         &       ...          &       ...           &       [ 0.038]$^d$   &       [ 0.038]$^d$  &    0.038   \\                                                                                                                                                                                                                                                                                                                                                                                                                      
$\phi_{K}    $ /Hz                          &       ...          &       ...           &       [ 0.372]$^d$   &       [ 0.372]$^d$  &    0.372   \\                                                                                                                                                                                                                                                                                                                                                                                                                      
$|\mu_a|     $ / D                          &       ...          &       ...           &        ...                        &         ...             &    2.75             \\                                                                                                                                                                                                                                                                                                                                                                                            
$|\mu_b|     $ / D                          &       ...          &       ...           &        ...                        &         ...             &    0.76             \\                                                                                                                                                                                                                                                                                                                                                                                            
$N$$^e$                                     &        34          &        9            &       130                         &         164             &    ...              \\                                                                                                                                                                                                                                                                                                                                                                                                         
$J_{\text{max}}/K_{\text{max}} $            &        7/2         &        4/1          &       50/22                       &         50/22           &    ...              \\                                                                                                                                                                                                                                                                                                                                                                                            
$\sigma_{\text{fit}}$$^f$/ kHz              &        0.8         &        1.8          &       20.3                        &         18.1            &    ...              \\                                                                                                                                                                                                                                                                                                                                                                                                         
$\sigma_{\text{w}}$$^g$                     &        0.80        &        0.89         &       0.68                        &         0.71            &    ...              \\      
\hline
\hline
\end{tabular}
\end{center}
\begin{tablenotes}
      \small
      \item \textit{Notes}. $^a$ The numbers in parentheses are the parameter uncertainties in units of the last decimal digit.  $^b$ Rotational constants and dipole moment components were calculated in this work while the centrifugal distortion constants come from \cite{cabezas_discovery_2025}. \revt{The new ab initio values are obtained at the CCSD(T)-F12A/cc-pVTZ-F12 while the old ones at the CCSD/cc-pVTZ one, see Section \ref{sec:spec_qc} for more details. } $^c$ Fixed to the values derived for the main isotopologue. $^d$ Values in brackets have been fixed to the calculated ones. $^e$ Number of distinct frequency lines. $^f$Root mean square deviation of the fit. $^g$ Unitless (weighted) deviation of the fit.\\
    \end{tablenotes}
\end{table*}

\begin{figure*}
\centering
\includegraphics[angle=0,width=\textwidth]{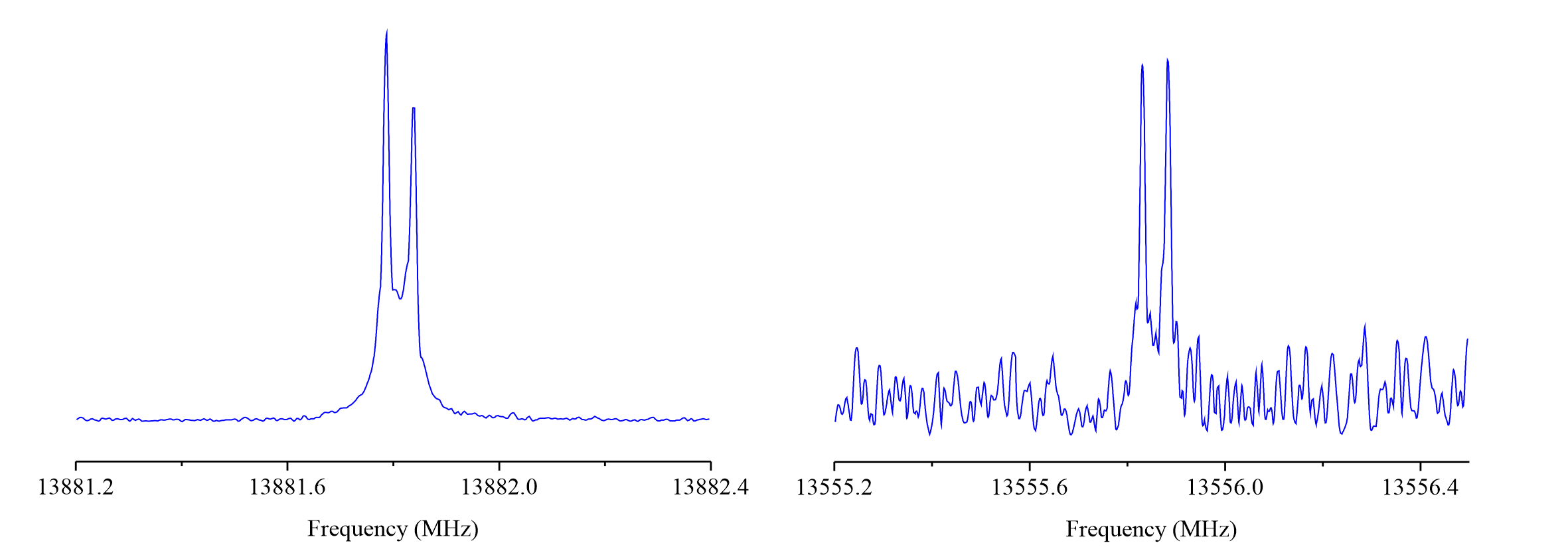}
\caption{Section of the FTMW spectrum of \syn\ propenethial showing 2$_{0,2}$-1$_{0,1}$ pure rotational transition of the main isotopologue (left) and the same transition for the $^{34}$S isotopologue (right). The spectrum for the main isotopologue was achieved by 100-shots of accumulation and that for the $^{34}$S isotopologue was measured by 1000-shots of accumulation. The coaxial arrangement of the adiabatic expansion and the resonator axis produces an instrumental Doppler doubling. \revt{The molecular resonant frequencies} are calculated as the average of the two Doppler components.} \label{fig:ftmw}
\end{figure*}

\begin{figure*}
\centering
\includegraphics[angle=0,width=\textwidth]{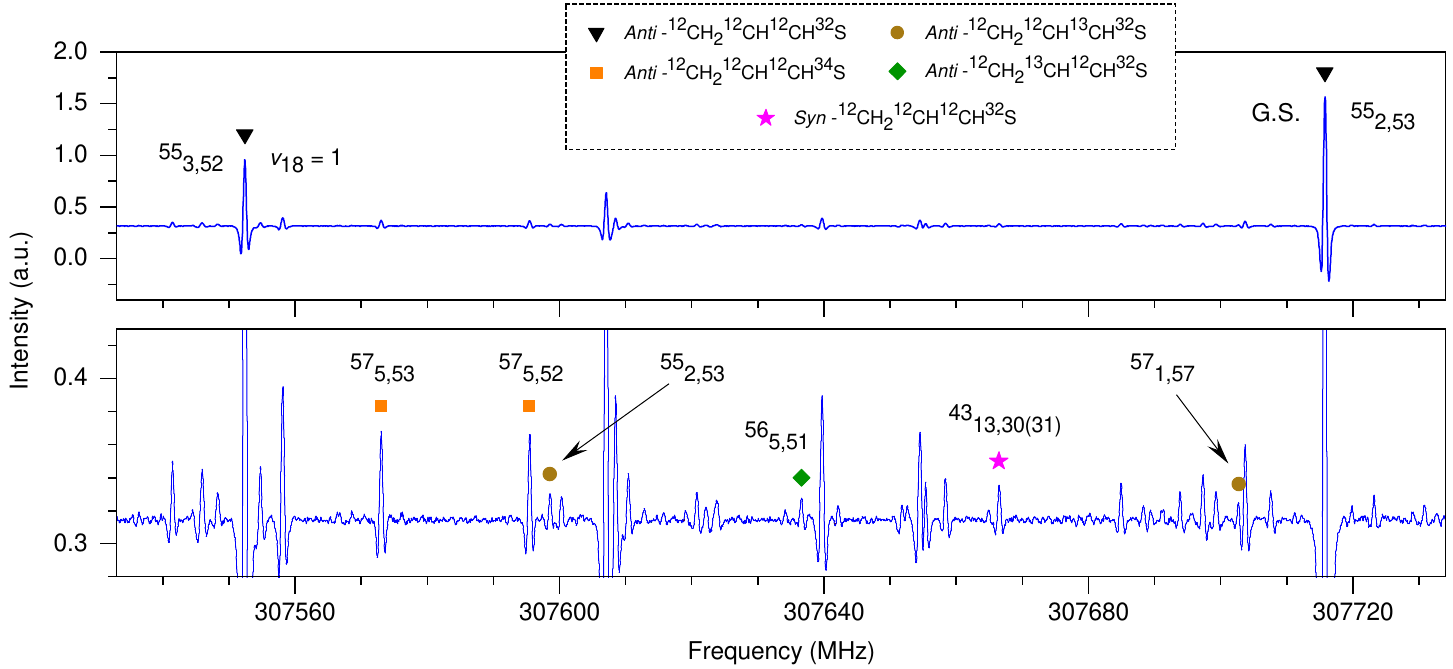}
\caption{Section of the mmw rotational spectrum of propenethial showing an example of a  rotational transition of the \syn\ isomer among the lines of the \anti\ isomer (ground state and $v_{18}=1$ excited vibrational state of the parent secies and the ground states of $^{13}$C and $^{34}$S isotopic species). Only the lower state quantum numbers for selected $a$-type R-branch transitions $J_{K_a,K_c}$ are indicated for clarity. The top panel shows the total intensity scale while the bottom panel shows a zoom of the total intensity.} \label{mmw}
\end{figure*}

Prior to the search for the rotational transitions of the \syn\ isomer, we checked the production of the \anti\ isomer of propenethial in the discharged jet. We observed very strong signals for some $a$-type R-branch rotational transitions,
indicating its efficient production. Then we searched for the $K_a=0,1$ $a$-type R-branch transitions of \syn\ isomer predicted in the 21~GHz region. A deep scan on this frequency region allowed us to detect some lines attributable to the target species. Further scans at higher frequencies allowed us to observe and assign additional $a$-type R-branch rotational transitions to the \syn\ isomer of propenethial. An initial fit allowed us to obtain a set of experimental molecular constants and in this manner accurately predict $K_a=2$ $a$-type R-branch transitions as well as $b$-type R-branch and Q-branch transitions. The final dataset consists of 34 rotational $a$- and $b$-type transitions with quantum numbers $0\leq J\leq7$ and $0\leq K_a\leq 2$, see an example of these lines in Fig.~\ref{fig:ftmw}. All the lines were analyzed with the \textsc{Spfit} program \citep{Pickett1991}, using a Watson’s A-reduced Hamiltonian for asymmetric top molecules \citep{Watson1977}. The analysis rendered the experimental rotational and quartic centrifugal distortion constants listed in the first column of Table \ref{constants}. As it can be seen, the experimental data for the \syn\ isomer, both the rotational and centrifugal distortion constants, agree very well with the theoretical values. However, to confirm the identity of the detected species, we searched also for the $^{34}$S isotopologue of the \syn\ isomer in its natural abundance (see Fig. \ref{fig:ftmw}). Its rotational transitions appear at the expected frequencies considering the isotopic shift for $^{34}$S. Only $a$-type R-branch transitions were observed, a total of seven lines, and their analysis led to the spectroscopic constants presented in the second column of  Table~\ref{constants}.

Having assigned the \syn\ isomer transitions in the microwave region, it was logical to extend the analysis to the millimeter-wave (mmw) region by searching for its transitions in our previously recorded spectrum. Locating its rotational features solely on the basis of the ab initio predicted molecular constants proved extremely difficult due to the spectral congestion and the expected weakness of the \syn\ isomer lines. It turned out, that its lines reach intensities similar to those for the $^{13}$C isotopic species of the \anti\ isomer (see Figure \ref{mmw}). The availability of experimental spectroscopic constants from the FTMW measurements thus proved invaluable for the reliable identification of the \syn\ isomer transitions in the 144--179 and 293--321~GHz frequency regions. A total of 130 rotational lines were measured for the \syn\ isomer of propenethial in the mmw region, allowing us to extend the range of the quantum numbers to $19\leq J\leq50$ and $0\leq K_a\leq 22$. The analysis of the data was carried out in the same manner as for the FTMW measurements and resulted in improved values of the quartic centrifugal distortion constants, as well as the determination of two sextic centrifugal distortion constants, $\Phi_{JK}$ and $\Phi_{KJ}$. The remaining sextic centrifugal distortion constants were kept fixed to those from ab initio calculations. The third column of Table~\ref{constants} lists the spectroscopic constants obtained using only the data set from the mmw frequency region, while the fourth column presents the results of a global fit that combines both the FTMW and the mmw measurements. The latter represents the most precise set of spectroscopic constants as it benefits from the complementary strengths of the two data sets -- the highly accurate $a$-type and $b$-type transitions provided by the FTMW measurements and the high‑$J$ and high‑$K_a$ transitions available in the mmw region.

Taking into account the observed spectral intensities of several rotational transitions, the relative abundance of each isomer under both experimental conditions can be estimated. For the FTMW experiment we found an approximate population ratio \anti:\syn\~=~4.5:1. formation occurs \rev{around} 1110-1200~K, in accordance with previous results using the same experimental setup \citep{Cabezas2016b}. This value is estimated assuming that the interconversion between the \syn\ and \anti\ forms is not likely due to the energy barrier height. In addition we accept that each isomer is cooled down separately by collisions in the supersonic expansion with the nascent production ratio almost \rev{unchanged}. On the other hand, in the mmw experiments we found an approximate population ratio \anti:\syn\~=~83:1 which is in agreement with the relative energy of \syn\ isomer and an equilibrium temperature of 298~K.\revt{We note the slight difference between the ratios obtained in the discharge and the thermal equilibrium constant derived in Section \ref{sec:discussion}. The former does not include thermal contributions (vibrational entropy for example) that may be irrelevant under discharge conditions, whereas the latter does.}

\subsection{Observations} \label{sec:observations_results}

\begin{figure*}
\centering
\includegraphics[width=0.95\linewidth]{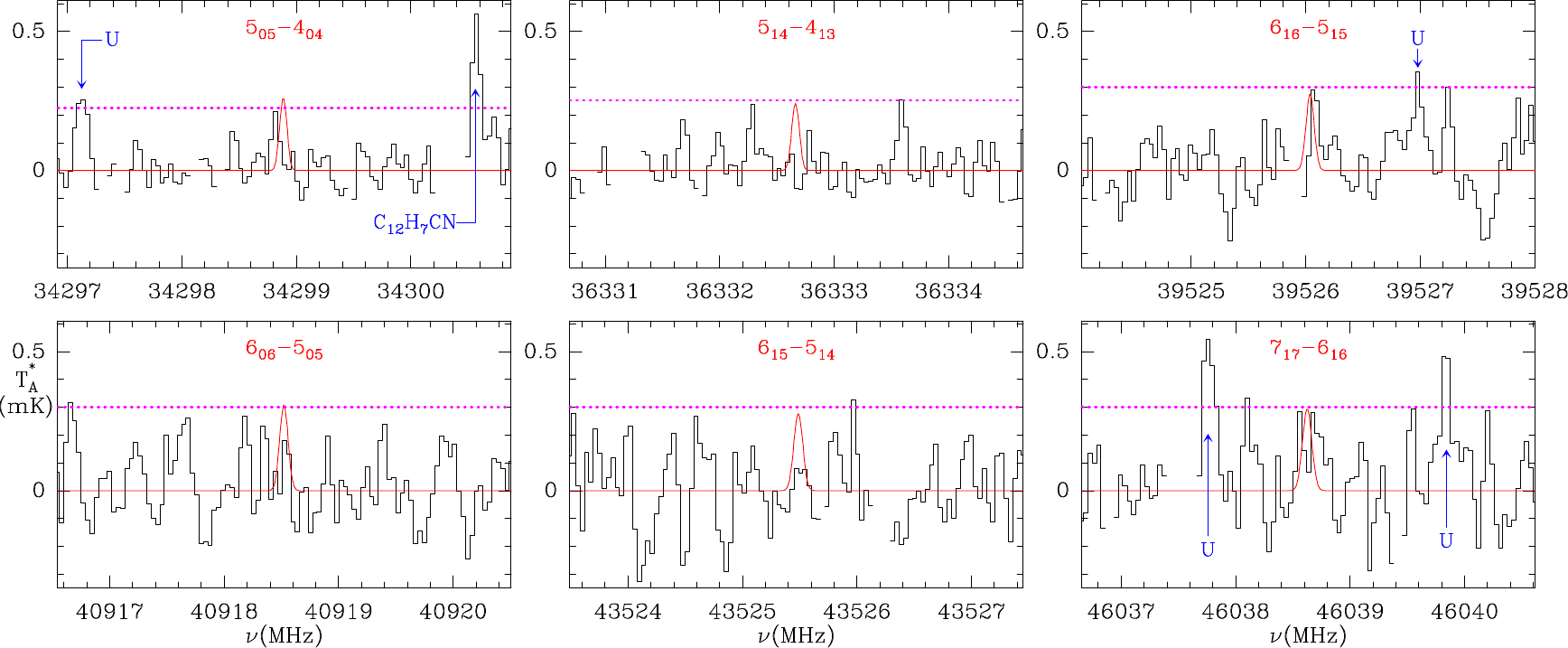}
\caption{Spectra of TMC-1 in the Q band at the frequencies of the most favorable transitions of \syn-\ce{CH2CHCHS}. Black histograms show the observed spectra while red lines correspond to the computed synthetic spectra for a column density of 1.5\,$\times$\,10$^{10}$ cm$^{-2}$. The dotted pink line represent the 3$\sigma$ level. The abscissa corresponds to the rest frequency, assuming a local standard of rest velocity of 5.83 km s$^{-1}$. The ordinate is antenna temperature in \revt{millikelvin}.}
\label{fig:lines}
\end{figure*}

In order to search for \syn-\ce{CH2CHCHS} in the \textsc{Quijote} data of TMC-1 we calculated the line intensities assuming local thermodynamic equilibrium (LTE) and adopting a rotational temperature of 9 K, the gas kinetic temperature in TMC-1 \citep{Agundez2023}. The most intense lines expected in the Q band are $a$-type transitions with $K_a$\,=\,0,1 (see Fig.\ref{fig:lines}). We did not detect any of them and therefore used the data to derive an upper limit to the column density of \syn-\ce{CH2CHCHS}. To derive this upper limit, once the physical parameters were fixed in \textsc{Madex}\footnote{\url{https://nanocosmos.iff.csic.es/madex} \citep{2012EAS....58..251C}} code, we varied the column density of the species until the modelled
intensities reach the 3 sigma noise level of the data for each transition line. We did not allow the model fit to be greater than any observed line. We again assumed LTE adopting a rotational temperature of 9 K, a line width of 0.6 km s$^{-1}$, and an emission size of 80$''$ of diameter, typical parameters in TMC-1 \citep{Agundez2023,Cernicharo2023}. The most stringent upper limit is imposed by the non detection of the $5_{1,4}$-4$_{1,3}$ line at 36332 MHz (see Fig.\,\ref{fig:lines}). We derive a 3\,$\sigma$ upper limit to the column density of \syn-\ce{CH2CHCHS} in TMC-1 of 1.5\,$\times$\,10$^{10}$ cm$^{-2}$. The derived upper limit shows a small dependency on the
assumed rotational temperature. For example, for Trot=6 K
the upper limit should be multiplied by 1.1. Taking into account that the column density of \anti-\ce{CH2CHCHS} in TMC-1 is 4.4\,$\times$\,10$^{10}$ cm$^{-2}$ \citep{cabezas_discovery_2025}, the \syn-to-\anti\ column density ratio for propenethial is $<$\,0.34.

\subsection{Selectivity from gas-phase reactions} \label{sec:gas-phase-reactions}

\begin{figure}
    \centering
    \includegraphics[width=\linewidth]{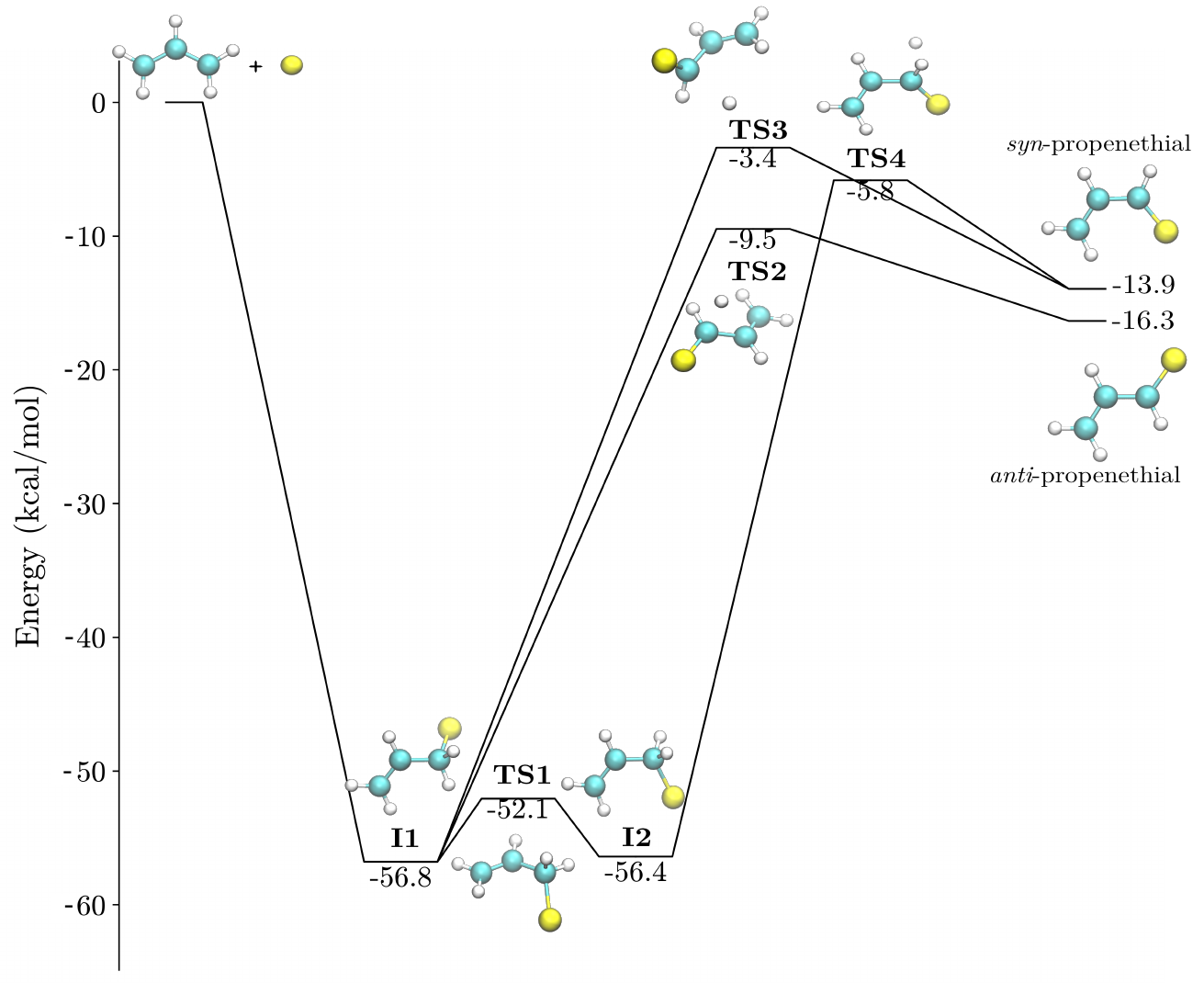}
    \caption{\revt{Simplified potential energy diagram for the shortest path in the formation of \syn\ and \anti\ propenethial.}}
    \label{fig:pes}
\end{figure}

In Figure \ref{fig:pes} we show the reduced potential energy surface for the shortest path in the gas-phase formation of propenethial. The reaction in our model proceeds as follows. First, the sulphur atom is captured in \textbf{I1} with a well energy of -56.8 \kcalmol. The conformational arrangement of \textbf{I1} can be described as a \textit{gauche} one with two non-equivalent H elimination positions, one leading to \anti\ propenethial (\textbf{TS2}) and another one leading to \syn\ propenethial through \textbf{TS3}. Both elimination transition states are rather high in energy, but clearly separated by approximatedly 6 \kcalmol (-9.5 against -3.4 \kcalmol). Additionally to the direct H-elimination, our model system includes the possibility of isomerizing from \textbf{I1} to a second intermediate \textbf{I2} that belongs to the $C_{s}$ symmetry point group, through \textbf{TS1} through a small barrier of approximatedly 5 \kcalmol. The H-elimination from \textbf{I2}, owing to the symmetry plane, leads exclusively to the \syn\ isomer with a barrier of $\sim$50 \kcalmol. 

\begin{figure}
    \centering
    \includegraphics[width=0.5\linewidth]{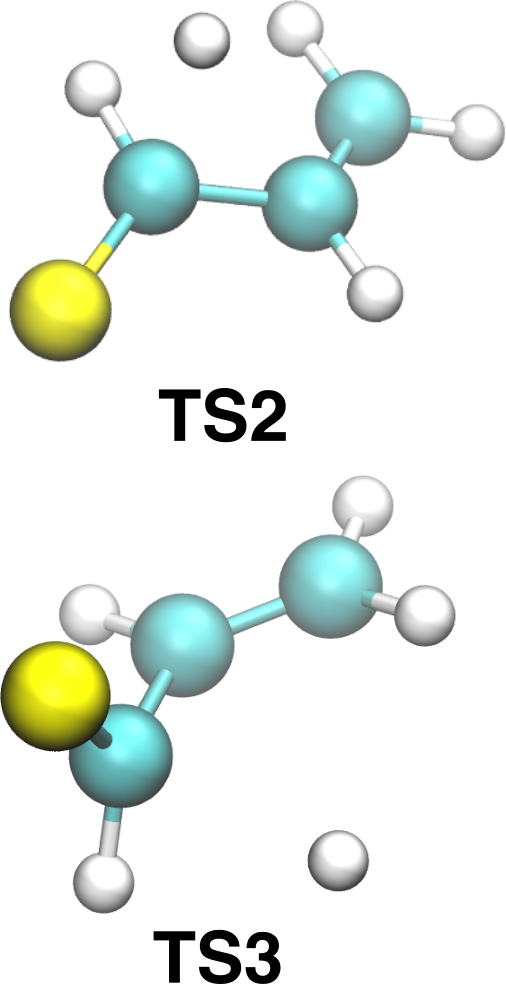}
    \caption{Larger representations of \textbf{TS2} and \textbf{TS3}, see also Figure \ref{fig:pes}.}
    \label{fig:ts}
\end{figure}

Analyzing the reduced PES in Figure \ref{fig:pes}, we observe that elimination from \textbf{I1} is the most favorable channel. This trend can be rationalized in terms of steric interactions at the transition state. To quantitatively support this interpretation, which can already be qualitatively visualized in Figure \ref{fig:ts}, we performed a local energy decomposition (LED) analysis of the transition states using a domain local pair natural orbital (DLPNO) wavefunction and the same basis set employed throughout this work \citep{RN102,RN257,RN188,RN165,RN66}. \revt{The results of the LED analysis are shown in Appendix \ref{sec:appA}, where the small deviations with Figure \ref{fig:pes} originate from the lack of ZPVE in the LED analysis and the use of a local treatment of the electron correlation}. The origin of the increased activation energy leading to the \syn\ isomer can be traced back to the interaction between the departing H atom and the neighboring hydrogen atom of the \ce{CH2} group in \textbf{I1}. Such an interaction is absent in \textbf{TS2}, whereas it is clearly present in \textbf{TS3}. Consistently, the LED analysis shows that the electrostatic contribution between the H atom and the \ce{CH2CHCHS} fragment, which dominates steric repulsion, is 15.5 \kcalmol\ lower in \textbf{TS2} than in \textbf{TS3}. The exchange contribution is also larger in \textbf{TS2} than in \textbf{TS3}, although the difference is significantly smaller (2.1 \kcalmol). In contrast, \revt{the departing H} intrafragment energy is more stabilizing in \textbf{TS3} by 13.9 \kcalmol. After inclusion of the remaining, less significant LED terms, these effects combine to yield an overall energy difference of approximately 6 \kcalmol\ between \textbf{TS2} and \textbf{TS3}. This analysis provides a clear physical interpretation of why the corresponding stereoisomer is not efficiently formed in the ISM. Similar arguments may apply to other undetected or previously overlooked species whose formation is controlled by subtle steric effects at the transition state.


After analysing the electronic factors behind the favored formation of a given propenethial isomer, we present the branching ratios for the reaction, obtained as described in Section~\ref{sec:gas}. The branching ratios are found to be in a nearly constant 15.8 \anti-\syn\ ratio in the 40--200 K range, revealing a very significant prevalence of the \anti\ isomer from the \ce{S + C3H5} reaction, in agreement with the conclusions drawn from the energetic analysis in the previous paragraphs. At 40~K, the lowest temperature considered in our master equation calculations (with the trend extrapolated down to 10~K), the formation of the \anti\ isomer accounts for 94.1 \% of the total production. At higher temperatures, the ratio does not significantly vary, reaching 92.7 \% at 200 K. This provides an elegant explanation for why \syn\ propenethial remains undetected in TMC-1, considering that the \anti\ isomer was already not very abundant. It is reasonable to argue that other routes could contribute to the formation of \syn-propenethial, and that our reaction modelling might be simplified with respect to the overall formation network of this molecule, \rev{like ion-molecule or other alternative neutral-neutral reaction. However, we indicate that our exploration of the \ce{C3H5S} PES, revealed that only the reaction considered here was barrierless. Additionally grain chemistry can play a role.} In \citet{cabezas_discovery_2025} we also discussed routes on dust grains based on non-thermal processes. Our initial exploration of these routes has indicated a lack of \rev{selectivity toward conformers}, and therefore \syn\ propenethial could be formed through them. However, branching ratios are much more difficult to determine for these reactions (see, for example, \citealt{sanz-novo_conformational_2025}). The grain route is nevertheless expected to be secondary with respect to the gas-phase route, owing to its non-thermal nature combined with the scarce availability of reactants for non-thermal mechanisms. 

\rev{The gas-phase, \ce{C3H5 + S}, reaction involves atomic sulphur and \ce{C3H5}, a resonance-stabilised radical. In analogy with \ce{CH2CCH} \citep{Agundez2021a}, \ce{C3H5} is expected to be abundant owing to its enhanced stability. However, the very low dipole moment of \ce{C3H5} (\ce{CH2CHCH2}, the lowest energy isomer), 0.05 D \citep{crabtree_ab_2025} makes the detection very challenging}. \rev{Furthermore, the reaction network of \ce{C3H5} is currently less well constrained than that of \ce{CH2CCH}. Nevertheless, preliminary astrochemical models \revt{(Briefly shown in Appendix \ref{sec:appB})} that incorporate part of the \ce{CH2CCH} chemistry into the \ce{C3H5} network suggest that the two radicals have comparable abundances, although both remain severely underpredicted.}

Alternative gas-phase routes either proceed through kinetic barriers when the reactants are small and abundant \citep{cabezas_discovery_2025}, or should produce propenethial in much lower quantities as a secondary product for reactions involving larger fragments. Therefore, by constraining the branching ratio of the \ce{C3H5 + S} reaction, we can estimate a theoretical upper bound for the \syn\ isomer based on the observational abundance of the \anti\ isomer, which already lies at the lower end of abundances in the TMC-1 inventory.  By multiplying the column density of \anti\ propenethial ($4.4\times10^{10}$~cm$^{-2}$) by the ratio of branching fractions at 40~K, we obtain a theoretical column density of $2.8\times10^{9}$~cm$^{-2}$, which is consistent with the observational upper limit (Section~\ref{sec:observations_results}), but found to be about one order of magnitude lower.


\subsection{\anti-\syn\ conversion at 10 K} \label{sec:tunneling}

\begin{figure}
    \centering
    \includegraphics[width=\linewidth]{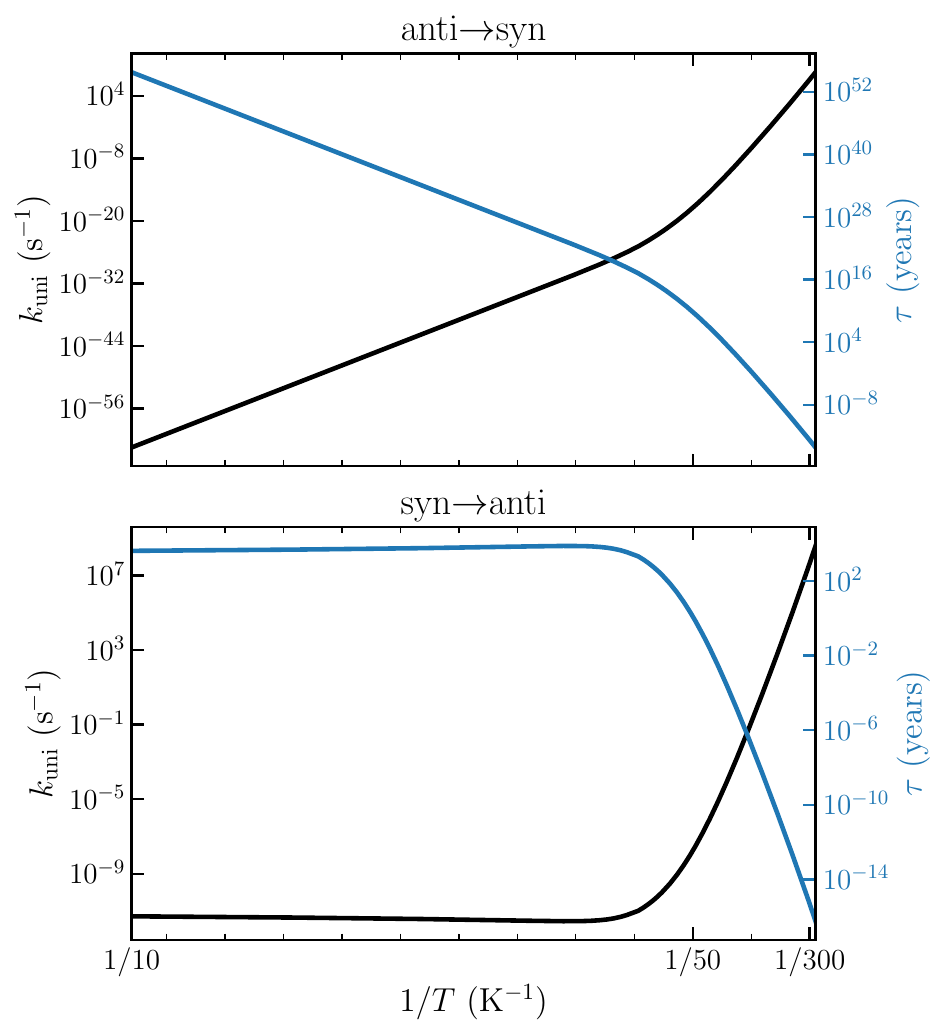}
    \caption{Forward (upper panel) and backward (bottom panel) rate constants for unimolecular isomerization (k$_{\rm uni}$ in inverse seconds) and lifetimes ($\tau$ in years).}
    \label{fig:tau}
\end{figure}

The energy gap between \anti\ and \syn\ \ce{CH2CHCHS} is 2.4 \kcalmol (1207 K) at the CCSD(T)-F12/cc-pVTZ-F12//SCS-MP2/aug-cc-pVTZ level. The torsional barrier, on the other hand, is 7.7 \kcalmol (3873 K) from the \anti\ isomer (endothermic) and 5.3 \kcalmol (2666 K) from the \syn\ one. In both cases, the activation energy is too high for the reaction to proceed in either direction, despite the extremely displaced chemical equilibrium, with $\mathcal{K_{\rm anti-syn}}$=7.1$\times$10$^{-53}$. We derived unimolecular rate constants, $k_{\rm uni, syn \rightarrow anti}$(10 K)=5.3$\times$10$^{-12}$ s$^{-1}$, and, $k_{\rm uni, anti \rightarrow syn}$(10 K)=3.7$\times$10$^{-64}$ s$^{-1}$, representing the latter an \rev{infinitely} slow process. The rate constants and associated lifetimes as a function of temperature are portrayed in Figure \ref{fig:tau}. The lifetimes in years for the \anti\ isomer clearly indicate the practical impossibility of transforming \anti\ into the \syn\ isomer. In contrast, the lifetimes of the \syn\ isomer at low temperatures are on the order of $10^{3}$ years, thanks to the increased efficiency of the reaction in the exothermic direction, facilitated by quantum tunneling. Such timescales are not particularly long on astronomical scales and could allow for a \syn$\rightarrow$\anti\ conversion. This would make the detection of \syn\ isomers (explicitly conformers in this context) in astrochemical environments unlikely if the production rate of the \syn\ isomer is slower than its conversion rate. 

\section{Discussion} \label{sec:discussion}

The spectroscopic characterization of \syn\ propenethial and subsequent astronomical search for it in TMC-1 has lead to a non-detection of this isomer. From observations, the upper limit to the column density that we derive is 1.5$\times$10$^{10}$ cm$^{-2}$. Our theoretical characterization on the other hand, places the upper limit nearly one order of magnitude lower, at 2.8$\times$10$^{9}$ cm$^{-2}$, only considering the optimistic scenario for its formation shown in Section~\ref{sec:gas-phase-reactions}. We certainly did not consider all the possible formation routes for \syn\ propenethial, including grain-surface routes that would certainly form it, but the here presented chemical reaction route is likely one of the dominant ones in the TMC-1 environment, where a prevalence of gas phase formation routes is expected. Besides, should \syn\ stereoespecific route be dominant, the calculations in Section \ref{sec:tunneling} and Figure \ref{fig:tau} reveal that the \syn\ isomer would be converted into the \anti\ one in a timescale of approximately $10^{3}$ years, meaning that, in order to have appreciable amounts of \syn\ propenethial, its formation rate should be higher than its conversion rate. 

\begin{figure}
    \centering
    \includegraphics[width=\linewidth]{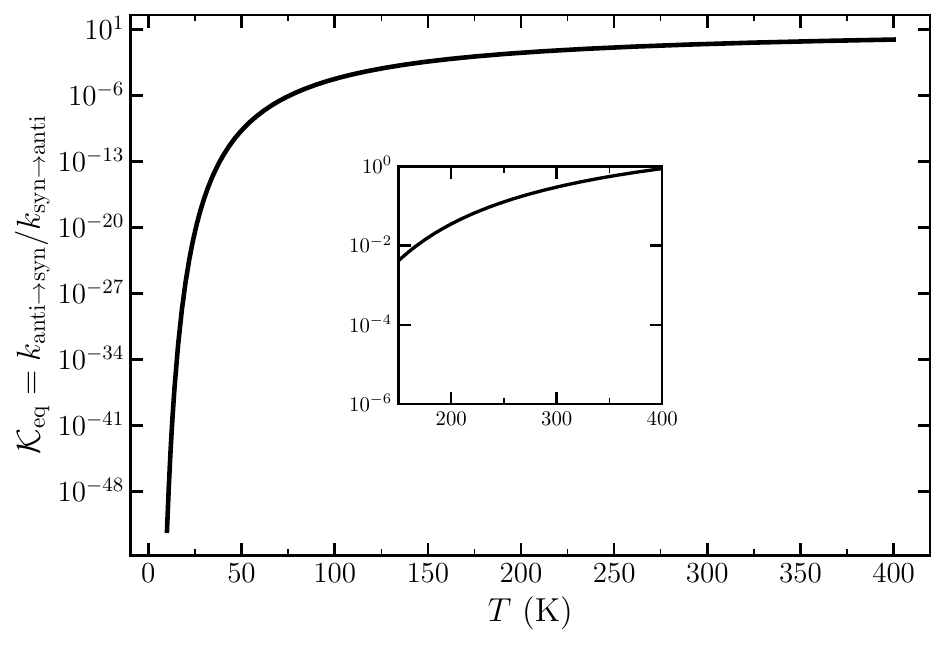}
    \caption{Equilibrium constants, $\mathcal{K}$, defined in the \anti $\rightarrow$ \syn direction as a function of temperature. The inset is a zoom in the 150--400 K region.}
    \label{fig:K}
\end{figure}

Although propenethial has not been detected outside TMC-1, it is meritorious to determine what would be the expected detectability for \syn\ \ce{CH2CHCSH}. From the detailed balance of the \anti\ and \syn\ forward and backward rate constants we can obtain the equilibrium constant $\mathcal{K}$ at different temperatures. The information provided by $\mathcal{K}$ speaks exclusively about equilibrium, without explicit reference to kinetics that were covered in the Section \ref{sec:tunneling}.  These values of $\mathcal{K}$ are shown in Figure \ref{fig:K}. While tiny at low temperatures, its value reaches 10$^{-2}$ at 150 K, reaching almost parity at 400 K, owing to the small energy difference between the two isomers. \rev{In fact, the \ce{CH2CHCSH} conformer pair exhibits a rather unusual combination of relative thermodynamic stability and interconversion barrier among interstellar conformational isomers (see, e.g., the conformational isomers listed in Table 1 of \citealt{rivilla_interstellar_2026}). The energy difference between the two conformers is sufficiently large to yield negligible equilibrium constants at 10 K, while the interconversion barrier remains low enough to permit conformational exchange over long timescales at low temperatures. } 

In warm molecular clouds and hot cores, however, the two species can be closer to equilibrium. We overall expect that \syn\ propenethial will not be a relevant species in the ISM. Finally, it is also important to mention, that all the here derived conclusions \rev{pertain to} interstellar environments shielded from intense UV fields. Although only confirmed for \textit{cis}-\textit{trans} formic acid \citep{cuadrado_trans-cis_2016}, the presence of low-lying electronic states for a pair of isomers can induce rapid isomerization, clearly affecting the potential detectability of the high-energy isomer of the pair, a process known as molecular photoswitching.

Concerning the larger picture of \ce{CH2CHCHS} in the sulphur chemistry of dark clouds and its S/O ratios, we can place the here presented results in the context left in our study on the detection of \ce{CH3CHS} \citep{2025A&A...693L..20A}. There, we indicate that, when moving H$_{x}$C$_y$S for higher values of $y$, a clear decrease in the abundance of the S-bearing species with respect to the O-bearing analogs is observed, with the S/O ratio going from being dominant in \ce{C2S} to be as low as $\sim$0.02 for \ce{CH3SH}/\ce{CH3OH} and \ce{CH3CHS}/\ce{CH3CHO}. For \ce{CH2CHCHS} on the contrary, the S/O ratio seems to flatten, with a value of approximately 0.2 \citep{cabezas_discovery_2025}. It is not clear whether \ce{CH2CHCHS} actually flattens the S/O ratio or if, by contrast, \ce{CH3CHS}/\ce{CH3CHO} and \ce{CH3SH}/\ce{CH3OH} are rarities of the overall trend. In either case, the S/O ratio for sulphur bearing COMs will be markedly dependent on the dominant formation, and specially destruction routes of the species. For example, it is known that \ce{CH3CHO} is a resilient molecule \citep{molpeceres_hydrogenation_2025}, a trait also attributable to \ce{CH3OH}, specially in comparison with \ce{CH3SH} \citep{2023ApJ...944..219N}, where the dominant hydrogenation route is found to be \ce{CH3 + H2S}. The study of the destruction routes of sulphur bearing COMs, including \ce{CH3CSH} and \ce{CH2CHCHS} is a promising avenue to understand the S/O ratios for these species, something that is in our immediate plans. 

\section{Conclusions} \label{sec:conclusions}

In this work, we combined experimental, observational, and theoretical efforts to advance our understanding of both sulphur chemistry and isomeric processes in cold dark clouds. Although the detection of \ce{CH2CHCHS} proved unsuccessful, its non-detection allows us to further constrain the mechanisms that lead to isomeric imbalances in the ISM and provides key information on the inventory of organic sulphur in these environments.

Our investigation can be summarized as follows:

\begin{itemize}

    \item The two isomers of \ce{CH2CHCHS} are separated by merely 2.4 \kcalmol, calculated at the CCSD(T)-F12/cc-pVTZ-F12//SCS-MP2/aug-cc-pVTZ level of theory, with a calculated torsional barrier of 7.7 \kcalmol in the \anti-\syn~direction and 5.3 in the reverse one (exothermic).

    \item The combination of an electric discharge source with FTMW spectroscopy enables the observation of the rotational transitions of the high-energy \syn\ isomer of propenethial, which are also observed in the millimeter-wave spectrum.

    \item Using our newly acquired spectroscopic data, we searched for \syn-propenethial in TMC-1 as part of the ultradeep observations of the \textsc{Quijote} line survey. The search was unsuccessful, and we derive an upper limit to the column density of \syn-propenethial of $1.5 \times 10^{10}$ cm$^{-2}$.

    \item The non-detection of \syn-propenethial provides observational constraints on the relative abundance of sulphur-bearing aldehyde analogues in TMC-1, contributing to a more complete assessment of the sulphur budget in cold dark clouds.

    \item The absence of \syn-propenethial is rationalized through dedicated quantum chemical calculations addressing both tunneling spontaneous unimolecular interconversion and gas-phase reactivity.

    \item The study of the simplest gas-phase formation route highlighted in \citet{cabezas_discovery_2025} reveals a strong \rev{conformer selective route} towards the \anti-isomer of propenethial. A detailed analysis of the origin of this selectivity shows that steric hindrance at the transition state results in a higher-energy pathway from a \textit{gauche} intermediate leading to \syn-\ce{CH2CHCHS}, in contrast with the less hindered H-elimination pathway that yields \anti-\ce{CH2CHCHS}. Assuming the investigated reaction is the most important one at most times in a molecular cloud evolution, we derive a theoretical upper limit for the molecule even lower than the observational one, of 2.8$\times$10$^{9}$ cm$^{-2}$. 

    \item The calculated lifetimes of \syn-\ce{CH2CHCHS} indicate that, in the absence of external perturbations such as chemical reactions or photoprocesses, most of the \syn\ isomer would convert into the more stable \anti\ form within a few thousand years, facilitated by quantum tunneling. This implies that any viable formation route must compete with this relatively rapid interconversion in order to enable detection in cold environments, which appears unlikely given the chemical complexity of the system and the available inventory of TMC-1.
    
    \item \rev{Both chemical formation routes and tunneling-mediated interconversion support the thesis that \ce{CH2CHCHS} will be found predominantly in the \anti\ form in cold molecular clouds. However, this simplified rationalization still requires the construction of a complete chemical network for propenethial, as the current agreement relies on a single potentially important reaction among the many processes that may contribute to the chemistry of \ce{CH2CHCHS}. Most of these reactions remain unexplored, both theoretically and experimentally.}

    \item The calculated equilibrium constants for the \anti/\syn\ pair of \ce{CH2CHCHS} show that, under thermodynamic equilibrium conditions, which are not achieved at 10 K, the isomeric ratio becomes less extreme at temperatures above 150 K, where the \syn/\anti\ fraction reaches values of $10^{-2}$--$10^{-1}$. Warmer environments may therefore be more favorable for the detection of the high-energy isomer, although the lower-energy form has not yet been detected in such regions.

\end{itemize}

In summary, our combined experimental, observational, and theoretical study provides a coherent rationalization of the molecular inventory in the ISM, offering an elegant explanation for the absence of the high-energy isomer of \ce{CH2CHCHS}. Beyond the specific case addressed here, our results raise broader questions regarding the relationship between the geometrical arrangement of reactive intermediates in gas-phase processes and the resulting \rev{conformer selectivity} of the reaction. In particular, the finding that the \textit{gauche} isomer of the \ce{CH2CHCH2S} intermediate suppresses the formation of the \syn\ isomer of \ce{CH2CHCHS} highlights the decisive role of conformational constraints along the reaction pathway.

Whether similar conformational effects systematically govern isomeric outcomes in other astrochemically relevant systems remains an open question. Exploring this possibility may ultimately enable the formulation of predictive guidelines linking intermediate geometry to isomer-specific product distributions in interstellar chemistry. This represents a promising direction for future investigations.

\section*{Acknowledgements}

This work is supported by ERC grant No. 101218790 (\textsc{Isocosmos}) funded by the European Union. Views and opinions expressed are however those of the author(s) only and do not necessarily reflect those of the European Union or the European Research Council Executive Agency. Neither the European Union nor the granting authority can be held responsible for them. GM also acknowledges the support of the grant RYC2022-035442-I funded by MICIU/AEI/10.13039/501100011033 and ESF+ and from PID2024-156686NB-I00 and project 20245AT016 (Proyectos Intramurales CSIC). We also acknowledge funding support from Spanish Ministerio de Ciencia, Innovación y Universidades through grant PID2023-147545NB-I00. Part of the laboratory characterization was funded by the Czech Science Foundation (GACR, grant No. 24-12586S). L.K. gratefully acknowledges this financial support. G.E. acknowledges support from the Spanish grant PID2022-137980NB-I00, funded by MCIN/AEI/10.13039/501100011033/FEDER UE.

\section*{Data Availability}

\rev{A Zenodo repository including laboratory spectroscopic data, isomerization rate constants and supporting structures is stored at \url{https://doi.org/10.5281/zenodo.21628971}. Any other required data can be obtained upon request.}



\bibliographystyle{mnras}
\bibliography{./example} 




\appendix

\section{\revt{Local Energy Decomposition Contributions} }
\label{sec:appA}

\begin{table}
\caption{\revt{Relative LED contributions for \textbf{TS2} and \textbf{TS3}. All energies are in \kcalmol. Values are reported relative to the lower value of each LED contribution between TS2 and TS3. Small deviations with energies reported in Figure \ref{fig:pes} originate from the lack of ZPVE and the use of a local treatment of electron correlation in the analysis}}
\label{tab:led}
    \centering
\begin{tabular}{lcc}

\hline
Contribution & \textbf{TS2} & \textbf{TS3}\\
\hline
Intra (\ce{CH2CHCHS}) & 0.0 & 1.9\\
Intra (\ce{H})        & 13.9 & 0.0\\
Electrostatics        & 0.0 & 15.5\\
Exchange              & 0.0 & 2.1\\
Dispersion            & 0.2 & 0.0\\
Non-dispersion        & 0.0 & 0.2\\
Total                 & 0.0 & 5.6\\
\hline
\end{tabular}
\end{table}

\section{\revt{Preliminary chemical model for \ce{C3H5} abundance} }
\label{sec:appB}

\begin{figure}
    \centering
    \includegraphics[width=\linewidth]{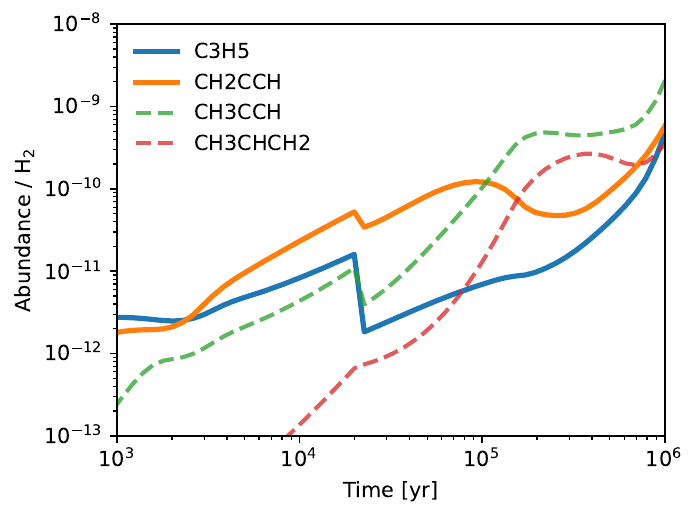}
    \caption{Exploratory chemical model to investigate the abundance of the \ce{C3H5} radical. All abundances are reported with respect the abundance of \ce{H2}}
    \label{fig:chemicalmodel}
\end{figure}

\revt{In the main text we indicate that preliminary gas-grain models indicate that \ce{C3H5} (the suggested precursor for \ce{CH2CHCHS}) should be an abundant molecule, but its intrinsically low dipole moment makes it really hard to detect. Our models are based on the Nautilus gas-grain code \citep{ruaud_gas_2016} using the KIDA reaction network for both gas and grain processes \citep{wakelam_kinetic_2012, 2024AA...689A..63W} and the default gas and grain parameters of a standard dark cloud modeled with this code. Only additions to the provided inputs include a series of gas-phase chemical processes that are considered in analogy with the chemistry of the \ce{C3H3} radical or recent investigations, namely:}

\begin{align}
    \ce{CH3CHCH2 + CN &-> C3H5 + HCN} \label{eq:one} \\
    \ce{CH3CHCH2 + C2H &-> C3H5 + C2H2} \label{eq:two} \\
    \ce{C + C2H6 &-> C3H5 + H } \label{eq:three}
\end{align}

\revt{Reaction \ref{eq:one} rate coefficient is taken from \citet{mallo_isomer-specific_2025}, in the case of reaction \ref{eq:two} we assume the same rate constant as the previous. Finally, reaction \ref{eq:three} is taken by analogy to the KIDA reaction:}
\begin{equation}
    \ce{C + C2H4 -> CH2CCH + H }
\end{equation}
\revt{with the same rate coefficient. Apart from the additions, we also deactivated the destruction reaction:}
\begin{equation}
    \ce{C3H5 + H -> CH3CCH + H2},
\end{equation}
\revt{also by analogy with the chemistry of \ce{CH2CCH}. The results of the chemical model are shown in Figure \ref{fig:chemicalmodel}. The model predicts that at evolutionary times of $\sim$2–5$\times$10$^{5}$ yr and beyond, comparable to the estimated age of TMC-1, the abundances of the two radicals become similar. We also note that the model significantly underestimates the absolute abundance of \ce{CH2CCH}, for which observational constraints are available. Assuming an \ce{H2} column density of 1$\times$10$^{22}$ cm$^{-2}$, the predicted \ce{CH2CCH} column density is of the order of a some 10$^{11}$ cm$^{-2}$, approximately two orders of magnitude lower than the observed value (8.7$\times$10$^{13}$ cm$^{-2}$, \citealt{Agundez2021a}). By analogy, we expect the model to similarly underestimate the abundance of \ce{C3H5}. This supports our assumption that \ce{C3H5} is an abundant molecule in TMC-1.}



\bsp	
\label{lastpage}
\end{document}